\documentclass[prb, twocolumn,showpacs]{revtex4}

\usepackage{graphicx}

\begin{document}
\title{Nature of the Quantum Phase Transition in Quantum Compass  Model}
\author{Han-Dong Chen}
\affiliation{Department of Physics, University of Illinois at Urbana-Champaign, Urbana, IL 61801}
          
\author{Chen Fang}
\affiliation{Department of Physics, Purdue University, West Lafayette, IN 47907}
\author{Jiangping Hu}
\affiliation{Department of Physics, Purdue University, West Lafayette, IN 47907}

 \author{Hong Yao}
\affiliation{Department of Physics, Stanford University, Stanford, CA 94305}

\begin{abstract}
In this work, we  show that the quantum compass model on an square lattice can be mapped to a fermionic model with local density interaction. We introduce a mean-field approximation where the most important fluctuations, those perpendicular to the ordering direction, are taken into account exactly. It is found that the quantum phase transition point at $J_x=J_z$ marks a first order phase transition. We also show that the mean field result is robust against the remaining fluctuation corrections   up to the second order.
\end{abstract}

\maketitle
\section{Introduction}
Quantum compass model \cite{Kugel1973} has recently attracted great interests \cite{Mishra2004,Doucot2005,Dorier2005,Nussinov2005}. It was originally proposed as a simplified model to describe some Mott insulators with orbit degeneracy described by a pseudospin. In particular, compass model describes a system where the anisotropy of the spin coupling is related to the orientation of bonds. More recently, it was also proposed as a realistic model to generate protected qubits \cite{Doucot2005}. It has been argued that the eigenstates of the quantum compass model are two and only two fold degenerate \cite{Doucot2005}. The two-fold degenerate ground state can be an implementation of a protected qubit if it is separated from the low energy excitations by a finite gap. Symmetry will protect weak noise destroying the degeneracy. Based on the protected qubits, which has high quality factor, a scalable and error-free schema of quantum computation can be designed\cite{SHOR1995,Preskill1998}. 
The two-fold degenerate state has been shown to be gapped from low energy excitations based on the results calculated in a small size system\cite{Doucot2005}. However, the results can not be extended to large size systems. The results from both spin wave study and exact diagonalization have suggested that the system develops a spontaneously symmetry-broken state in the thermodynamic limit\cite{Dorier2005}.
In the spin-wave study, the Hamiltonian is expanded around the uniform classical ground state up to the first order of $1/S$. It is found that there is a directional ordering of the ground state\cite{Dorier2005}.  From finite size diagonalization of samples with size up to $5\times 5$, it is shown that on clusters of dimension $L\times L$, the low-energy spectrum consists of $2^L$ states which collapse onto each other exponentially fast
with $L$. At the symmetric point of $J_x=J_z$, $2\times 2^L$ states collapse exponentially fast with $L$ onto the ground state. From both of the spin-wave analysis and exact diagonalization, a first order phase transition at the symmetric point seems most favorable. However, the spin-wave analysis ignores very important fluctuations while the exact diagonalization is limited by small sample size. In this work, we first show that the spin-$\frac{1}{2}$ quantum compass model on a square lattice can be exactly mapped to a fermionic model with local density interactions. 
Normally, after performing a Jordan-Wigner transformation, one expects a non-local gauge interaction between fermions in the fermionic Hamiltonian\cite{Fradkin1989}. However, due to the special structure of both the spin interactions and the lattice, we show that the gauge interaction for the compass model is absent, which allows us to apply conventional approximation techniques developed for  electron systems to analyze the original spin model. Our approximation method automatically takes into account the most important fluctuations, those perpendicular to the ordering direction.
The remaining fluctuations can also be studied in the perturbative approach. It is shown that our conclusion is robust against the perturbative corrections. Our results support that a first order phase transition happens at $J_x= J_z$ between two  different states with  spin ordering  along either $x$ or $z$ directions.   

\section{Compass model and fermionization}

The compass model appears rather simple on the first look,
\begin{eqnarray}
  H=-J_x \sum_{i,j}S^x_{i,j}S^x_{i+1,j}- J_z \sum_{i,j} S^z_{i,j} S^z_{i,j+1},
 \label{H-full}
\end{eqnarray}
where $(i,j)$ is a two dimensional coordinate. The sign of $J_x$ and $J_z$ is not important. We can always introduce a transformation to bring them into $J_x,J_z\geq0$. We assume temperature $T=0$ throughout this work and set $\hbar=1$ for simplicity.

This model has some interesting symmetries, which have been discussed in details in literature \cite{Doucot2005,Nussinov2005}. We first recall two types of symmetry generators of this model\cite{Doucot2005}
\begin{eqnarray}
  P_i=\prod_j 2S^x_{i,j},\quad
  Q_j=\prod_i 2S^z_{i,j}. \label{symmetry-PQ}
\end{eqnarray}
 It has been shown that the symmetry of this model leads to a one-dimension type of behavior and directional ordering \cite{Doucot2005,Nussinov2005,Mishra2004}. It is also shown \cite{Nussinov2005} that this model is dual to recently studied models of $p+ip$ superconducting arrays\cite{Xu2004,Xu2005}, which also show the effect of dimensional reduction \cite{Xu2005}.

With the help of the symmetry operators defined in Eq.(\ref{symmetry-PQ}), we find that it is possible to fermionize the full compass model. The resulted fermionic model has BCS-type pairing along one direction and nearest neighboring interactions along the other direction. The Jordan-Winger transformation can be defined as
\begin{eqnarray}
S^+_{i,j}&=&\left(\prod_{j'<j}Q_{j'}\right)
\left(\prod_{i'<i}\left[2c^\dag_{i',j}c^{}_{i',j}-1\right]\right)c^\dag_{i,j}\\
S^z_{i,j}&=&c^\dag_{i,j}c_{i,j}^{}-\frac{1}{2},
\end{eqnarray}
where $c_{i,j}^{}$ annihilates a fermion at site $(i,j)$. 
The original compass model Eq.(\ref{H-full}) is transformed into
\begin{eqnarray}
H&=&-\sum_{i,j}\left[J_z n_{i,j}^{}n_{i,j+1}^{}-J_z n_{i,j}^{}\right.
\nonumber\\
&&\left.+ \frac{J_x}{4}\left(c^{}_{i,j}-c^\dag_{i,j}\right)
\left(c^{}_{i+1,j}+c^{\dag}_{i+1,j}\right)
\right] 
\end{eqnarray}
up to a constant. Because the interaction along $z$ direction is quartic, the string-type interaction or gauge interaction disappears.

\section{Phase diagram}
\subsection{General considerations}

The symmetries defined in Eq.(\ref{symmetry-PQ}) also have profound implications on the possible ordering of this model.
Based on the fact that $[P_i,P_{i'}]=[Q_j,Q_{j'}]=[P_i,Q_jQ_{j'}]=[P_iP_{i'},Q_j]=0$ and $[P_i,Q_j]\neq 0$, it is found that each eigenstate of this model is two and only two fold degenerate \cite{Doucot2005}. This leads to the observation that any finite system cannot have an ordering that is characterized by some finite mean-field expectation value of either or both $\langle S^x_{i,j}\rangle$ and $\langle S^z_{i,j}\rangle$. For instance, let us assume the ground state $|\Omega\rangle$ of a $L\times L$ system has finite $\langle S^z_{i,j}\rangle$. One can thus apply $P_i$ onto $|\Omega\rangle$ and obtain $2^L$ degenerate ground state, which is of course inconsistent with the above result of double-degeneracy. However,  ordering is still possible when the system goes to the thermodynamic limit where spontaneous symmetry breaking happens. The excitation gap that separates the true ground state and other $2^{L-1}$ low energy excitation states collapses exponentially as the system size goes to infinity\cite{Dorier2005}. In this case, the spontaneously broken symmetries are the $Z_2$ symmetries of the one-dimensional Ising chain along the ordering direction. Let us consider a system $L_x\times L_z$ and let both $L_x$ and $L_z$ increase to infinity. If the ordering is along the $z$ direction, one should observe $2^{L_x-1}$ states collapse onto the true ground state and the gap vanishes exponentially, namely $e^{-L_z/L_0}$ with some length scale $L_0$. Since the spontaneously broken symmetries are a large number of copies of discrete symmetry $Z_2$, there is no corresponding Goldstein mode. Furthermore, one can also effectively view this spontaneously symmetry breaking as the breaking of a set of local $Z_2$ symmetries in a one-dimensional problem, which is along $x$ if the ordering is along $z$. This does not violates the Elitzur's theorem \cite{Elitzur1975} since in this case the effective local $Z_2$ symmetry is realized by a infinitely long Ising chain and thus the energy barrier to restore the symmetry is still infinite.

We shall now consider the phase diagram in the thermodynamic limit. We start from two extreme cases which are trivial to solve. We then study how the system evolves from one extreme to the other. It turns out that by studying this question it is possible to obtain the phase diagram as we shall show below.

\begin{figure}[t]
  \includegraphics{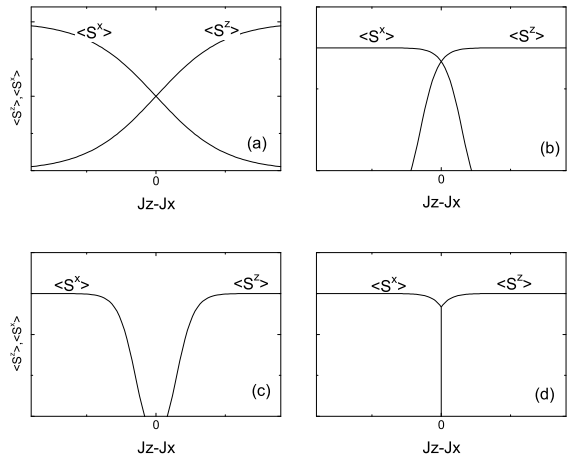}
  \caption{Four possible scenarios of the evolution of spin ordering. (a) Crossover scenario. (b) Two second order phase transitions with coexisting state in between. (c) Two second order phase transitions with disordered state in between. (d) First order phase transition scenario.}
\end{figure}

  Let us first start from the simplest limit $J_x=0$, where the model reduces to decoupled Ising chains. The ground state is magnetically ordered such that $\langle S^z\rangle = \frac{1}{2}$ up to an overall flip of all spins of each Ising chain. This ordered state thus has $2^L$ degeneracy for a $L\times L$ system. Similarly, at the limit of $J_z=0$, the spins are ordered along $x$ direction. We are thus interested in how the spins rotate from $z$ direction to $x$ direction as the ratio $J_z/J_x$ changes from $\infty$ to $0$ and what is the possible ground state at $J_x=J_z$. In very general, there are four possible scenarios. The first possibility is that there is no phase transition. The spins rotate continuously from $z$ to $x$ through a crossover. The second scenario corresponds to two second order phase transitions with coexisting of ordering along both $z$ and $x$ directions.  The third possibility is two second order phase transitions with disordered state in between. The fourth one is that the spins suddenly change from $x$ direction to $z$ direction, i.e., a first order phase transition happens at $J_z=J_x$.

\subsection{Mean-field phase diagram}
To find out which of the four possible scenarios happens, we start with limit $J_x=0$, where the system is ordered along $z$ direction, and increase $J_x$ (or decrease the ratio $J=J_z/J_x$). As $J_x$ is increased, the flipping of two spins on adjacent chains is introduced. This process tends to reduce the ordered moment along the $z$-direction. On the other hand, the ordering along $z$ direction acts as an effective transverse field and suppresses the ordering along $x$ direction of the spins on adjacent chains. This effect is described by a 1D Ising chain with a transverse field, which has a critical transverse field \cite{Pfeuty1970}.  When $J_x$ is much smaller than $J_z$, the ordering of spins along $x$ direction is fully suppressed by the adjacent spins ordered along $z$ direction. We then ask at what value of $J_x$ the spins start to order along the $x$ direction. For the crossover case, $\langle S_x\rangle$ has a finite value as long as $J_x$ is not zero. For the second scenario, two second order phase transitions with intermediate coexisting state, $\langle S_x \rangle$ starts to develop at some value of $J$ that is larger than $1$. For the last two cases, the expectation value of $S_x$ is zero for all $J_x<J_z$, while $\langle S_z\rangle$ takes a finite value for all $J_z>J_x$ in the first order phase transition scenario. To determine the behavior of $\langle S^x\rangle$ and $\langle S^z\rangle$, we self-consistently study the mean-field decoupled Hamiltonian 
\begin{eqnarray}
    H_{MF}/J_x&=&-\sum_{i,j} S^x_{i,j} S^x_{i+1,j}
    -\sum_{i,j} B_{eff}(i) S^z_{i,j},\label{H}\\
    B_{eff}(i)&=&J\left[\langle S^z_{i,j-1}\rangle
  +\langle S^z_{i,j+2}\rangle\right].
\end{eqnarray}
The mean-field decoupling is justified, since we start from the situation where $S^z$ is ordered and want to know when $S^x$ starts to develop a finite expectation value. This mean field decoupled Hamiltonian describes the most important fluctuation in the state with spins ordered along $z$ direction. The ordering along $x$ direction is suppressed by the ordering along $z$ direction, which behaves as an effective transverse field for $S_x$. On the other hand, the feedback effect on $S_z$ due to $S_x$ fluctuations is also included in this approach through the self-consistent condition
\begin{eqnarray}
    \langle S^z_{i,j} \rangle = \langle \psi |S^z_{i,j}|\psi\rangle,
\end{eqnarray}
where $|\psi\rangle$ is the ground state of the mean-field Hamiltonian $H_{MF}$.
To minimize the ground state energy, it is favorable to have 
$\langle S^z_{i,j}\rangle$ $j$-independent, {\it i.e.}, the moment on the same chain is uniform such that the $J_z$ terms are minimized. 
The mean-field Hamiltonian thus describes a series of decoupled Ising chains with an effective transverse magnetic field $B_{eff}=2J\langle S^z_{i,j}\rangle$ that is determined self-consistently.

To minimize the coupling energy along $z$-direction, one would expect $|B_{eff}(i)|=B_{eff}$ is $i$-independent. If $B_{eff}(i)=-B_{eff}$, we can rotate the spins on $i$-th row by an angle $\pi$ around the $x$-axis. Without losing generality, we can thus consider the case where $B_{eff}(i)=B_{eff}$.

In the Fermionic language, the mean-field decoupling introduced in Eq.(\ref{H}) corresponds to the decoupling of the interaction term $J_z n_{i,j}^{}n_{i,j+1}^{}-J_z n_{i,j}^{}$ into an effective chemical potential term
$\mu n_{ij}$ with
\begin{eqnarray}
\mu=2J\langle n_{ij} \rangle-J=B_{eff}.
\end{eqnarray}
This Hamiltonian can be diagonalized using Bogoliubov transformation
\begin{eqnarray}
  \gamma^{}_k = u_k c^{}_k + v_k c^\dag_{-k}~~~~~~
  \gamma^{}_{-k}=v_k c^\dag_k - u_k c^{}_{-k},
\end{eqnarray}
where $c_k$ is Fourier transformation of $c(i)$.
The $u_k$ and $v_k$ can be solved to yield 
\begin{eqnarray}
    u_k &=& \sqrt{\frac{1}{2}\left[1+\frac{(\cos k)/2-\mu}{E_k}\right]}\\
    v_k &=& i\sqrt{\frac{1}{2}\left[1-\frac{(\cos k)/2-\mu}{E_k}\right]}
\end{eqnarray}
with $E_k=\sqrt{1/4-\mu\cos k + \mu^2}$.
It follows that the expectation value of $S^z$ is
\begin{eqnarray}
\langle S^z\rangle =
\frac{1}{2\pi}\int_{0}^{2\pi}dk\frac{2B_{eff}-\cos k}{8E_k/J_x}\label{Sz}
\end{eqnarray}
The self-consistent equation thus reads
\begin{eqnarray}
   B_{eff}=-\frac{J}{\pi}
   \int_{0}^{2\pi}dk\frac{\cos k-2B_{eff}}{\sqrt{1-4 B_{eff}\cos k + 4 B_{eff}^2}}.\label{Eq-selfconsistent}
\end{eqnarray}

\begin{figure}[t]
  \includegraphics[width=2.4in]{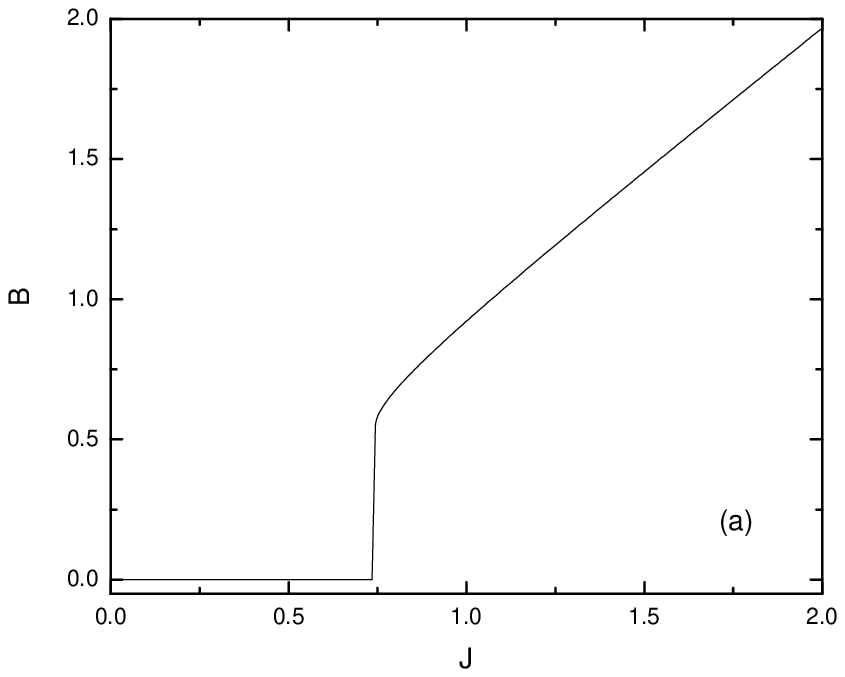}  \\ 
  \includegraphics[width=2.4in]{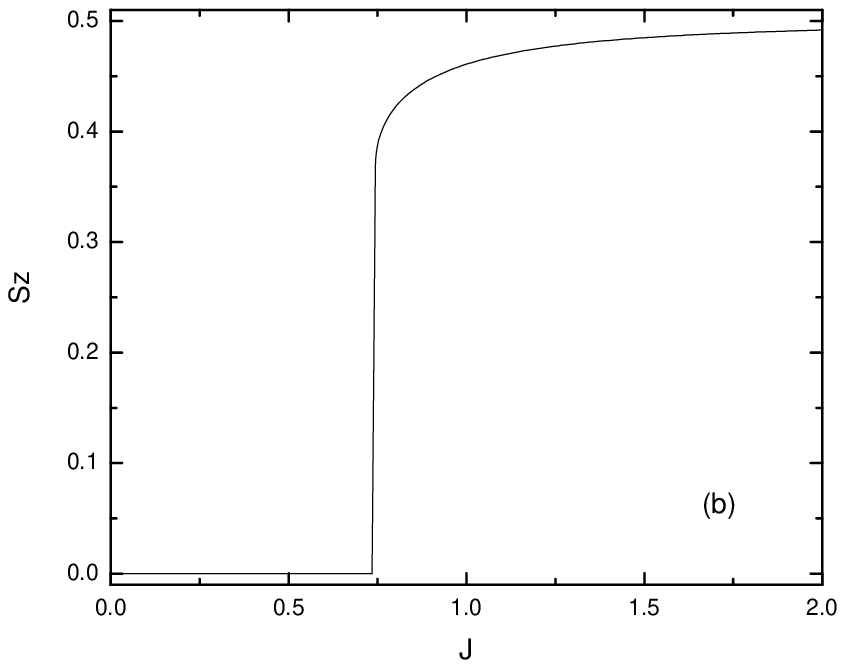}  
  \caption{The solution to self-consistent equation (\ref{Eq-selfconsistent}). There is no non-trivial solution for $J<0.7446$. (a) The effective transverse field $B_{eff}$ vs $J$. (b) The ordered moment along $z$ direction $\langle S^z\rangle$ vs $J$.}\label{FIG-solution}
\end{figure}

From the results shown in Fig.\ref{FIG-solution}, we know there are nontrivial solutions to the self-consistent equation (\ref{Eq-selfconsistent}) for $J\geq 0.7446$. The jump happens at $B_{eff}\approx 0.65>B_c=1/2$. Below $B_c$, which is the critical point of the Ising chain in a transverse magnetic field, $S_x$ starts to develop a nonzero expectation value. We notice two important results that can help us determine the phase diagram. Firstly, the critical point $B_c$ happens at a $J$ smaller than the symmetry point $J=1$, below which ordering along $z$ direction is no longer a reasonable assumption. For $J<1$, the ordering along $x$ direction is instead more favorable. This implies that $\langle S^x\rangle$ cannot develop a finite value for $J_x<J_z$. The first two scenarios, namely crossover scenario and two second-order phase transitions scenario, are thus excluded. Secondly, the self-consistent equation has non-trivial solutions for $J>0.7446$, which means spins order along $z$ direction for $J_z>J_x$. Combining these two points, we conclude that there is a first order phase transition at $J=1$. At $J>1$, the spins are ordered along $z$ direction while they are along $x$ direction for $J<1$. At $J=1$, the spins are either along $x$ or $z$ direction since our argument above shows that the ordering at $J=1$ along $x$ ($z$) direction induces a strong enough transverse field to suppress the ordering along $z$ ($x$) direction. This conclusion is consistent with the spin-wave analysis\cite{Dorier2005}. In spin wave analysis, fluctuations of both directions are taken into account partially. In our approach, the most important fluctuations, namely the ordering direction of weaker bond, is solved exactly.

\subsection{Beyond mean-field approximation}


Although the mean-field approximation has included the most important fluctuations, there is still concerns about the effect of the ignored fluctuations.
To further study the effect of fluctuations, we calculate the energy gap by perturbation. The perturbation comes from the interaction terms that are ignored in the mean-field Hamiltonian Eq.(\ref{H})
\begin{eqnarray}
H_{int}=-\sum_{ij}J_z n_{i,j} n_{i,j+1}+2J_z\langle n\rangle\sum_{ij}n_{i,j}.\label{Eq-interaction}
\end{eqnarray}

\begin{figure}[h]
\includegraphics{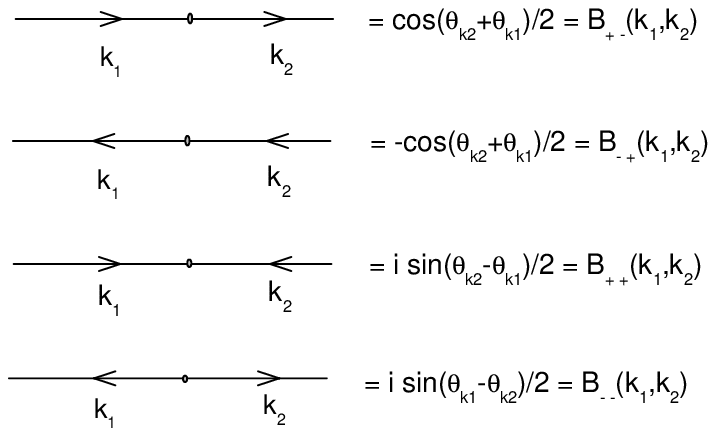}
\caption{Four possible contributions of $j$-th line to the vertex.}\label{FIG-Bcoeff}
\end{figure}

In terms of Feynman diagrams, there are hence two kinds of vertices: two-particle interaction vertex and external field interaction.
Let us first work out the Feynman rules for the two-particle interactions.
We observe that the interactions are between particles on nearest neighboring lines. The vertex of two-particle interaction thus consists of three parts, the coupling strength, the contribution from $j$-th line and the contribution from the neighboring $(j+1)$-th or $(j-1)$-th line. There are four possible types of contributions, which are given in Fig.\ref{FIG-Bcoeff} with 
\begin{eqnarray}
\tan \theta_k=\frac{-\sin k}{2 J_z/J_x+\cos k}.
\end{eqnarray}

\begin{figure}
\includegraphics{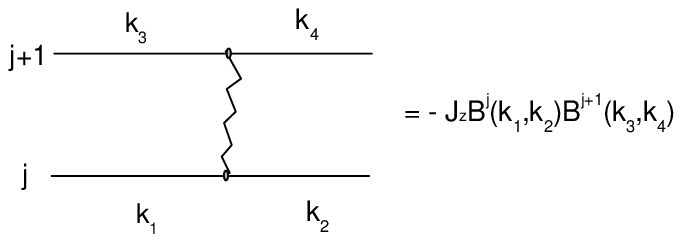}
\caption{The vertex of two-particle interaction.}\label{FIG-vertex}
\end{figure}

\begin{figure}
\includegraphics{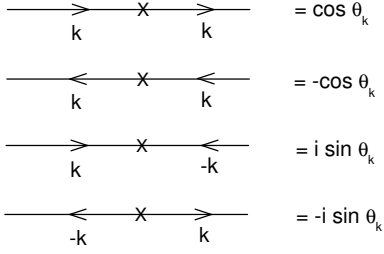}
\caption{The vertex for the external field term.}\label{FIG-vertex-external}
\end{figure}

The first line corresponds to $\langle\gamma_{k_2}^{}\gamma^\dag_{k_2}\rangle \langle\gamma_{k_1}^{}\gamma_{k_1}^\dag\rangle$. In this section, we will use $\langle A\rangle$ to denote the expectation value of operator $A$ in the un-perturbed mean-field ground state.  This term can be obtained by contraction from either $\langle\gamma_{k_2}^{}(\gamma^\dag_{k_2} \gamma_{k_1}^{})\gamma_{k_1}^\dag\rangle$ or $\langle\gamma_{k_2}^{}(\gamma_{k_1}^{}\gamma^\dag_{k_2} )\gamma_{k_1}^\dag\rangle$. The coherence factor of $(\gamma^\dag_{k_2} \gamma_{k_1}^{})$ in the two-particle interaction is $\cos \frac{\theta_{k_1}}{2} \cos\frac{\theta_{k_2}}{2}$ while the one of $( \gamma_{k_1}^{}\gamma^\dag_{k_2})$
is $\sin \frac{\theta_{k_1}}{2} \sin\frac{\theta_{k_2}}{2}$. Taking into account of the sign difference between these two contractions, we find $B_{+-}(k_1,k_2)=\cos\frac{\theta_{k_1}+\theta_{k_2}}{2}$. Similarly, we get the results for other three possibilities. The resulted vertex for two-particle interaction is thus $-J_zB^j(k_1,k_2)B^{j+1}(k_3,k_4)$ with corresponding Feynman diagram sketched in Fig.\ref{FIG-vertex}. Here, we have not shown the arrows associated with $k_1,k_2,k_1+q, k_2-q$. These arrows determine the corresponding subscripts of $B^{j+1}$ and $B^j$. The corresponding momentum conservation law for the vertex is $\delta(\pm k_1  \pm k_2  \pm k_3  \pm k_4)$ where the signs of $k_1(k_3)$ and $k_2(k_4)$ are the first and second subscripts of $B^j(B^{j+1})$. For instance, the momentum conservation for $B^j_{+-}(k_1,k_2)B^{j+1}_{+-}(k_3,k_4)$ is $\delta(k_1+k_3-k_2-k_4)$.
Similarly, we can obtain the vertex for the external field term as illustrated in Fig.\ref{FIG-vertex-external}

\begin{figure}
\includegraphics{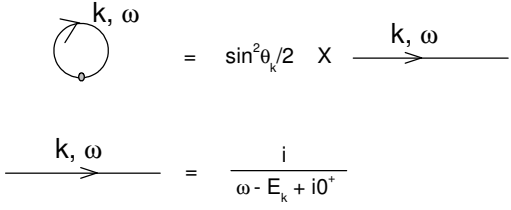}
\caption{The Feynman rules for bubble and propagator of particle.}\label{FIG-bubble}
\end{figure}

We are now at position to calculate the Feynman rule for bubbles. The clockwise bubble corresponds to $\langle \gamma^{}_k\gamma^\dag_k \rangle$ with coherence factor $\sin^2\frac{\theta_k}{2}$ while the counterclockwise one is $\langle \gamma^{\dag}_k\gamma^{}_k\rangle=0$ with coherence factor $\cos^2\frac{\theta_k}{2}$. The Feynman rules for bubble and propagator of particle is shown in Fig.\ref{FIG-bubble}
 A noticeable difference between this propagator and the one in
the familiar case of electrons under coulomb repulsion is that we
only have the plus sign before the infinitesimal imaginary part.
This is because in the current system, we don't have a Fermi sea as
ground state and therefore do not have a hole propagator. This
property of system largely reduces the number of diagrams we need to
calculate. Finally, every intermediate
4-momentum should be integrated with a factor of $(-i)/(2\pi)$.

\begin{figure}[t]
\includegraphics[width=3in]{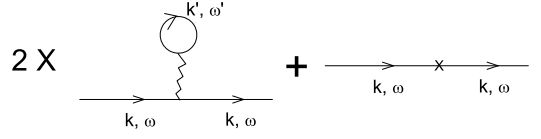}
\caption{First order correction to the self-energy.}\label{FIG-1st}
\end{figure}

\begin{figure}[t]
\includegraphics[width=3in]{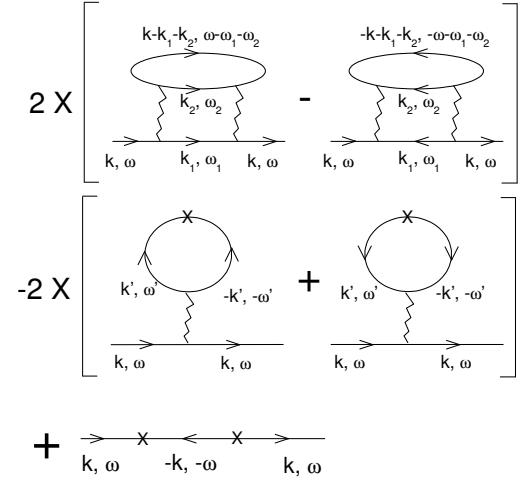}
\caption{Second order correction to the self-energy.}\label{FIG-second-order}
\end{figure}

With all the Feynman rules in hands, we are now ready to do perturbative calculations. The first order correction to the self-energy is given in Fig.\ref{FIG-1st}. The first term is the contribution from the interaction with nearest neighbor lines. The factor $2$ comes from the fact that there are two nearest neighbor lines. The second term is from the ``external field" term in Eq.(\ref{Eq-interaction}). The first order perturbation vanishes in the mean-field decoupling of Eq.(\ref{H}). The leading order correction is thus second order perturbation given by the connected amputated graphs as shown in Fig.\ref{FIG-second-order}. 

\begin{figure}[t]
\includegraphics[width=2.5in]{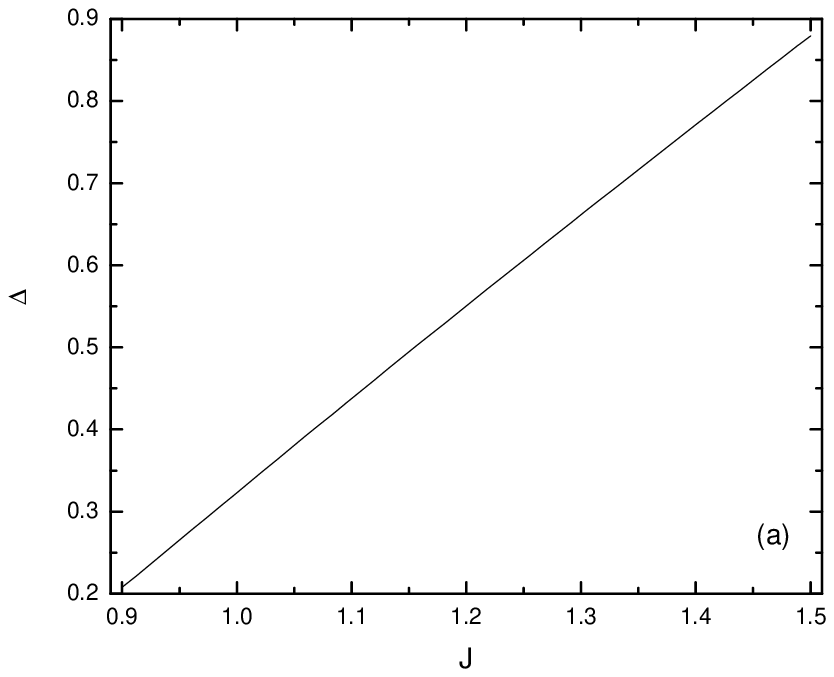}\\ 
\includegraphics[width=2.5in]{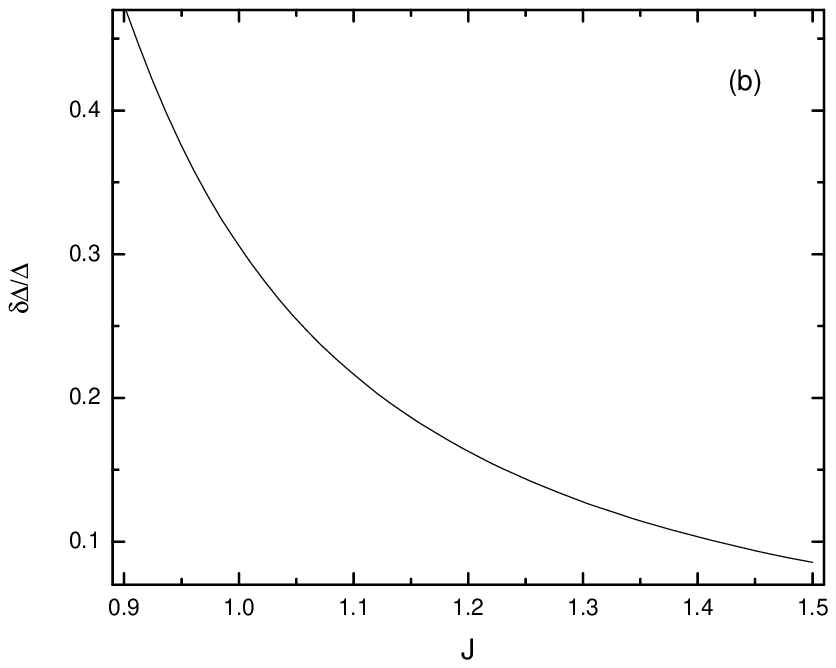}  
\caption{(a) Energy gap of the ground state with second order perturbation. (b) The ratio between second order correction and the energy gap itself.}\label{FIG-gap} 
\end{figure}

\begin{figure}[t]
\includegraphics[width=2.5in]{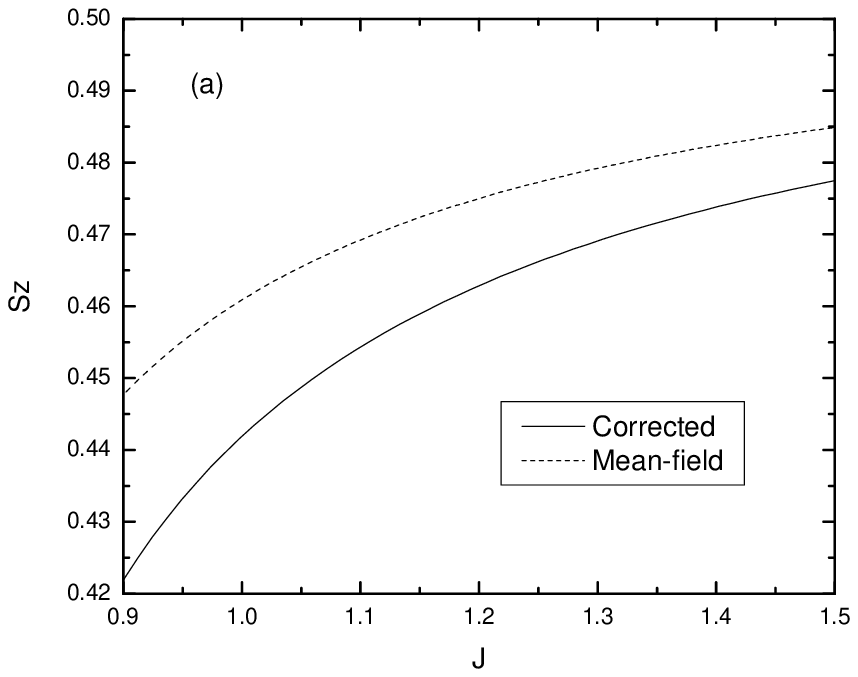} \\ 
\includegraphics[width=2.5in]{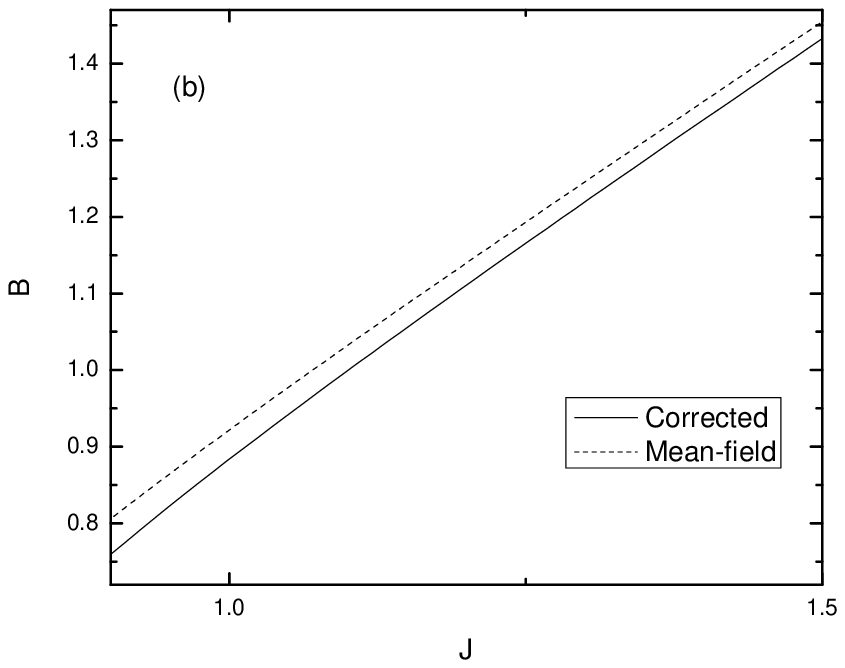} 
\caption{(a) The ordered moment along $z$ direction versus $J_z/J_x$. (b) The effective field $B_{eff}=2J_z S_z$ after second-order correction is included. Solid lines are the results with second order perturbation included while dash lines are mean-field results.}\label{FIG-Sz-J2}.
\end{figure}

There are five pieces of nonzero second order correction to the
single particle propagator. We detail the calculation of them
separately. The first two diagrams denote the self-energy process
involving two particle-particle interactions. The overall factor of
two stems from the fact that a particle on the, say, $j$th chain can
either interact with the $(j-1)$th or the $(j+1)$th chain. The
relative minus in the bracket is there because we invert three (an
odd number) of the propagators in the second diagram. Direct
application of the Feynman rules we derived yields
\begin{eqnarray}
I_1&=&\left(\frac{iJ_z}{2\pi}\right)^2\nonumber\\&\times&\int^{2\pi}_0
\frac{\cos^2\left(\frac{\theta_k+\theta_{k_1}}{2}\right) \sin^2\left(\frac{\theta_{k-k_1-k_2}-\theta_{k_2}}{2}\right)}
{\omega-\epsilon_{k_1}-\epsilon_{k_2}-\epsilon_{k-k_1-k_2}}
dk_1dk_2
\nonumber\\ ~ \\
I_2&=&\left(\frac{iJ_z}{2\pi}\right)^2 \nonumber\\ &\times&\int^{2\pi}_0
\frac{\sin^2\left(\frac{\theta_k-\theta_{k_1}}{2}\right) \sin^2\left(\frac{\theta_{-k-k_1-k_2}-\theta_{k_2}}{2}\right)}
{\omega+\epsilon_{k_1}+\epsilon_{k_2}+\epsilon_{k-k_1-k_2}}
dk_1d{k_2}, 
\nonumber\\
\end{eqnarray} 
where $I_1$ and $I_2$ are the contributions from the first and second pieces of Fig.\ref{FIG-second-order} respectively. The intermediate frequencies $\omega_1$ and $\omega_2$ have been integrated. 

The third and the fourth diagrams denote self-energy processes
involving one particle-particle interaction and one external field
perturbation. The factor of 2 stems from the same argument and the
overall negative sign can be obtained by working out the
contractions. Their analytical expressions $I_3$ and $I_4$ read 
\begin{eqnarray}  
I_3=I_4
=-2J_z\langle n\rangle
\left(\frac{iJ_z}{2\pi}\right) 
\int_{0}^{2\pi}\frac{\sin^2\theta_{k'}\cos\theta_{k}}
{2\epsilon_{k'}}dk',
\end{eqnarray} 
where again, the intermediate frequency has been integrated.

The last piece of the expression denotes process that involves two
external field perturbation. It is easy to obtain the analytical
expression as 
\begin{eqnarray}
I_5=-i(2J_z\langle n\rangle)^2~\frac{\sin^2\theta_k}{\omega+\epsilon_k}.
\end{eqnarray}

In this way, we have exhausted all possible self-energy processes to
the second order of $J_z$. The resulted sum of the diagrams is given by $-i\Sigma(\omega,k)=2I_1-2I_2-2I_3-2I_4+I_5$, where $\Sigma(\omega,k)$ is the
correction to the propagator of a quasi-particle carrying momentum $k$.
The energy of a particle can be defined as the singularity of its
propagator and readily obtained by solving the following equation
\begin{eqnarray} 
E(k)-\epsilon_k-\Sigma(E(k),k)=0.
\end{eqnarray}  
The energy gap is defined
as the lowest energy of excitation, which is numerically calculated
and shown in Fig.\ref{FIG-gap}. The second order correction amounts to about $30\%$ of the corrected energy gap itself at the symmetric point $J_x=J_z$.
In Fig.\ref{FIG-Sz-J2}, we also plot the ordered moment $S_z$ as a function of $J_z/J_x$. Again, the moment is reduced by a small second-order correction. At the symmetric point, the effective field is still larger than the critical value $B_c=1/2$ below which an ordered moment along $x$ direction becomes possible. We thus conclude that the mean-field results are robust against the second order perturbation.

\section{Conclusion}
In conclusion, we show that the compass model can be mapped to a fermion model with local density interaction. Through a self-consistent solution of the model, we argue that there is a first order phase transition at the symmetric point $J_x=J_z$. This conclusion is consistent with spin-wave analysis and recent numeric computations of the spectrum \cite{Dorier2005}. In our approach, the most important fluctuation, namely the fluctuations along the weak coupling direction, is taken into account exactly.  The fluctuations along the ordering direction is considered up to second order perturbation. It is shown that the result of our mean-field approximation is robust against such corrections.

We are grateful for the useful discussions with C. Xu, Prof. E. Fradkin, and Prof. A. J. Leggett.  This material is based upon work supported by the U.S. Department of Energy, Division of Materials Sciences under Award No. DEFG02-91ER45439, through the Frederick Seitz Materials Research Laboratory at the University of Illinois at Urbana-Champaign and  by the National Science Fundation, Division of Mathematical Physics, under 
Award NO.PHY-0603759.


\end{document}